\documentclass[runningheads]{llncs}
\usepackage{amsmath,amssymb,amsfonts}
\usepackage{algorithmic}
\usepackage{cite}
\usepackage{graphicx}
\usepackage{textcomp}
\usepackage{booktabs}
\usepackage{xcolor}
\usepackage{color, colortbl}
\usepackage{xurl,lineno,microtype}
\usepackage[hidelinks]{hyperref}

\def\BibTeX{{\rm B\kern-.05em{\sc i\kern-.025em b}\kern-.08em
    T\kern-.1667em\lower.7ex\hbox{E}\kern-.125emX}}

\newcommand{\bftab}{\fontseries{b}\selectfont}
\definecolor{Gray}{gray}{0.9}

\begin{document}

\title{Optimization of Image Acquisition for Earth Observation Satellites via Quantum Computing}

\titlerunning{Optimization of Image Acquisition for Earth Observation Satellites}

\author{Ant\'on Makarov\inst{1} \and
M\'arcio M. Taddei\inst{2} \and
Eneko Osaba\inst{3} \and \\
Giacomo Franceschetto\inst{2,4} \and
Esther Villar-Rodriguez\inst{3} \and
Izaskun Oregi\inst{3}
}

\authorrunning{Makarov et al.}

\institute{GMV, 28760, Madrid, Spain \and
ICFO - Institut de Ciencies Fotoniques, The Barcelona Institute of Science and Technology, 08860, Castelldefels, Barcelona, Spain \and
TECNALIA, Basque Research and Technology Alliance (BRTA), 48160 Derio, Spain \and
Dipartimento di Fisica e Astronomia “G. Galilei”, \\Università degli Studi di Padova, I-35131 Padua, Italy\\
\email{amakarov@gmv.com}}

\maketitle

\begin{abstract}
Satellite image acquisition scheduling is a problem that is omnipresent in the earth observation field; its goal is to find the optimal subset of images to be taken during a given orbit pass under a set of constraints. This problem, which can be modeled via combinatorial optimization, has been dealt with many times by the artificial intelligence and operations research communities. However, despite its inherent interest, it has been scarcely studied through the quantum computing paradigm. Taking this situation as motivation, we present in this paper two QUBO formulations for the problem, using different approaches to handle the non-trivial constraints. We compare the formulations experimentally over 20 problem instances using three quantum annealers currently available from D-Wave, as well as one of its hybrid solvers. Fourteen of the tested instances have been obtained from the well-known SPOT5 benchmark, while the remaining six have been generated ad-hoc for this study. Our results show that the formulation and the ancilla handling technique is crucial to solve the problem successfully. Finally, we also provide practical guidelines on the size limits of problem instances that can be realistically solved on current quantum computers.

\keywords{Quantum Computing  \and Satellite Image Acquisition \and Earth Observation \and Quantum Annealer \and D-Wave}
\end{abstract}

\section{Introduction}
The Satellite Image Acquisition Scheduling Problem (SIASP) \cite{bensana1999earth, wang2020agile} is of great importance in the space industry; every day, satellite operators must face the challenge of selecting the optimal subset of images to be taken during the next orbit pass of a satellite with respect to some notion of value from a set of requests made by their clients. For any given set of requests, there are constraints such as geographical proximity incompatibilities, on-board disk space availability or special image configuration requirements, which make it impossible to collect all the images. The problem becomes even harder to solve when we consider that, since requests arrive continuously over time, the planning must be updated several times a day, which requires the planning algorithm to be re-run many times over short periods of time. This fact makes execution speed crucial. Moreover, the industry is currently facing a myriad of challenges such as extending the problem to multiple satellites \cite{bianchessi2007heuristic, malladi2016satellite} or including climatic restrictions \cite{wang2018exact}, making it even more computationally expensive.

The SIASP and its extensions are extremely combinatorial in nature, which make them very challenging to solve with classical computing methods even for moderately sized instances. Traditionally, the problem has been tackled by exact integer linear programming algorithms based on branch-and-bound methods \cite{bensana1999earth, vasquez2003upper, ribeiro2010strong} or (meta-)heuristic and hybrid algorithms, which are usually preferred \cite{1399860, barkaoui2020new} since they are the only ones able to handle large scale problems and comply with execution-time limitations. 

On another note, Quantum Computing (QC) is emerging as a very promising alternative for dealing with optimization problems, where Quantum Annealing based computers such as the ones developed by D-Wave are especially suitable and might offer significant speedups in the future. The SIASP has been barely studied from the QC perspective, which makes this problem a great candidate to explore the possible near- to mid-term benefits that QC can bring to the table in the space industry.

Taking as motivation the scarcity of works conducted in this regard, our objective in this paper is to study SIASP from the QC perspective. To this end, we propose two distinct formulations to encode the variables of the problem.

Overall, 20 different problem instances have been used to test the adequacy of each formulation. On one hand, 14 of them have been directly taken from the well-known SPOT5 dataset \cite{bensana1999earth}, while the other six have been generated based on SPOT5 using an instance reductor implemented ad-hoc for this study. As solving alternatives, we assess the performance of four solvers available to the public from D-Wave, three purely quantum-annealing based and one hybrid. This allows us to assess the quality of our formulations as well as to test the limits of today's available hardware.

The rest of the paper is structured as follows. The following section offers some background on the problem dealt with in this paper. Next, Section \ref{sec:math} introduces the mathematical formulations employed for dealing with the problem by the quantum annealers considered. The description of the experimental setup is given in Section \ref{sec:exp}, along with the tests conducted and the discussion. Finally, Section \ref{sec:conclusions} ends the paper with several conclusions and future research following up from the reported findings.

\section{Background}\label{sec:Background}
QC is gaining significant notoriety in the scientific community, mainly due to its promising computational characteristics. It introduces new mechanisms, based on quantum mechanics, for solving a wide variety of problems more efficiently. This paradigm is expected to achieve a speed and/or precision advantage in modelling systems and in solving complex optimization problems, besides potential energetic benefits. More specifically, quantum computers have been employed in recent times for facing problems coming from diverse fields such as logistics \cite{osaba2022systematic}, economics \cite{orus2019quantum} or industry \cite{luckow2021quantum}.

Although the problem dealt with in this research is of great industrial interest, one of the main obstacles faced by any researcher when trying to tackle it through the quantum perspective is the shortage of previous work focused on this task. This fact implies a challenge to be faced and a research opportunity, which has motivated the realization of the present paper.

There is nevertheless a specific work, published in 2020 by Stollenwerk et al. \cite{stollenwerk2020image,stollenwerk2021agile}, which deals for the first time with the SIASP using the quantum paradigm. They investigate the potential and maturity of then-current quantum computers to solve real-world problems by carrying out an experimental study on a reduced number of small instances of the SIASP. To this end, they introduce a Quadratic Unconstrained Binary Optimization (QUBO) formulation, which served as inspiration for our research. In any case, since that formulation does not completely cover our needs, we have extended it in order to contemplate ternary constraints and multi-camera satellites.

The main contributions that our research provides with respect to that pioneering work can be summarized as follows:
\begin{itemize}
    \item The dataset employed in our research has been taken from realistic simulations of a satellite mission.
    \item The satellite considered has three cameras (instead of the single camera used in \cite{stollenwerk2020image}), two of which can be used in conjunction to take stereo images. This issue greatly increases the complexity of the problem, and gives rise to several possible formulations, of which two distinct ones are studied.
    \item In our paper, we consider ternary constraints, which require ancillary qubits to model them. We study two ways to encode those constraints and show the compressing power of one with respect to the other.
    \item We perform an experimental study on the performance of several quantum annealing based solvers.
\end{itemize}

Finally, it is interesting to mention the research recently published in \cite{zhang2018solving} and \cite{zhi2021variable}. These papers are focused on solving the SIASP using a quantum genetic algorithm. Despite the scientific contribution behind these investigations, they are outside the scope of the present paper since the methods developed fall within the area known as \textit{quantum-inspired evolutionary computation}. Thus, these techniques are really classical algorithms with the particularity of using concepts from quantum physics for their design.

\section{Mathematical Formulations of the Problem}\label{sec:math}
In this section, we focus on the mathematical formulation of the SIASP treated in this paper. First of all, we go in detail on the classical formulation of the problem. After that, the section gravitates around the quantum-oriented formulations chosen for the experimentation.

\subsection{Classical Formulation of the SIASP}\label{sec:classical}
Our classical formulation for the SPOT5 instances (we refer the reader to \cite{bensana1999earth} for a complete description of the structure of the instances) is largely inspired by \cite{vasquez2001logic}, and it can be stated in the language of mathematical programming as follows. Let $x_{i,j}$ be the binary decision variable, defined as:
\begin{equation*}
x_{i,j}=
    \begin{cases}
        1 & \text{if image $i$ is taken with camera $j$,}\\
        0 & \text{otherwise},
    \end{cases}
\end{equation*}
where $i \in \{1, 2, \ldots, N\}$ is the index representing the image requests, $N$ being the total amount of requests and $j \in \{1, 2, 3, 4\}$ the identifier of the camera. There are three physical cameras which can take mono images and we define camera 4 to represent the combined use of cameras 1 and 3 to take stereo images. Thus, the objective function to be optimized is:
\begin{align*}
\text{min} \quad & - \sum_i \sum_j w_i x_{i,j},
\end{align*}
where $w_i$ is the weight or value of taking image $i$. Note that although our task is to maximize the value, we can express it as the minimization of the negative of the objective. This optimization is subject to the following constraints:
\begin{subequations}
\begin{align}
\sum_j x_{i,j} &\leq 1 \quad \forall i \label{const:once} \\
x_{p, j_p} + x_{q, j_q} &\leq 1 \quad \forall \left((p, j_p),(q, j_q)\right) \in C_2\label{const:pairs}\\
x_{p, j_p} + x_{q, j_q} + x_{r, j_r} &\leq 2 \quad \forall \left((p, j_p),(q, j_q), (r, j_r)\right) \in C_3 \label{const:ternary} \\
x_{i,4} &= 0 \quad \forall i \in M \label{const:mono}\\
x_{i,j} &= 0 \quad \forall i \in S, \forall j \in \{1,2,3\} \label{const:stereo}\\
x_{i,j} &\in \{0, 1\} \label{const:binary}
\end{align}
where Constraint \eqref{const:once} forces the images to be taken only once. Constraint \eqref{const:pairs} represents the incompatibilities of taking certain pairs of images (set $C_2$) which arise from two images being too close geographically to each other to take both. Constraint \eqref{const:ternary} represents the ternary constraints (set $C_3$) related to the data flow restrictions of the satellite that do not allow to take more than two images at once. Constraints \eqref{const:mono} and \eqref{const:stereo} forbid mono images (set $M$) to be taken as stereo and stereo (set $S$) images to be taken as mono, respectively.
\label{eq:constraints}
\end{subequations}

\subsection{Formulations of the SIASP for its Quantum Solving}\label{sec:quantum}
In order to treat the problem with existing quantum-annealing hardware, we need a different formulation. Quantum annealers' QPUs are able to produce a final Hamiltonian of the form:
\begin{equation}
    H_F = \sum_i h_i\ Z_i + \sum_{i\neq j}J_{i,j} \ Z_i Z_j \ ,
\label{eq:generalIsingform}
\end{equation}
where $Z_i$ is the $z$ Pauli operator of the $i$-th qubit, $h_i$ is its on-site energy and $J_{i,j}$ the coupling between qubits $i,j$. This allows us to minimize the corresponding function on classical binary variables $x_i$ (obtained transforming $Z_i\to1-2x_i$). This is by analogy called \textit{energy}, and can be written as:
\begin{equation}
    E(\boldsymbol{x}) = \boldsymbol x^T Q \ \boldsymbol x \ ,
\label{eq:quboQmatrix}
\end{equation}
where $Q$ is a matrix that can be assumed symmetric without loss of generality. A QUBO consists in the minimization of such functions, and currently available quantum annealers such as D-Wave target this kind of problems specifically. The limitation to polynomials of order two comes from the fact that the Hamiltonian of Eq. \eqref{eq:generalIsingform} only couples qubits two by two, and is intrinsic to the hardware.

Additionally, the hardware is not able to couple every pair of qubits, hence an important feature of the QPU is its topology, i.e. the configuration of which qubits are in fact coupled, hence which off-diagonal terms of $Q$ are nonzero and can be tuned. For this study, three different QPUs have been used, which are \texttt{DW\_2000Q\_6} (\texttt{2000Q}), \texttt{Advantage\_system6.1} (\texttt{Advantage}) and \texttt{Advantage2\_prototype1.1} (\texttt{Advantage2}), have topologies called Chimera, Pegasus and Zephyr, respectively, which are increasingly interconnected \cite{DW-2000Q-Manual,DW-Adv61-Manual,DW-Adv2p11-Manual}. If the topology does not present all couplings needed for the problem, a typical solution is to embed a given qubit in more than one physical qubit. This is a source of overhead in the amount of qubits needed for a given instance. We make use of D-Wave's native algorithm based on minor embedding \cite{Choi2011}.

The transformation of our original problem into a QUBO formulation presents several challenges. Firstly, QUBO is unconstrained, i.e., we cannot explicitly impose Constraints \eqref{eq:constraints}. Secondly, currently available quantum hardware possesses few qubits, so reducing the number of variables is of utmost importance. Thirdly, it can only tackle problems written in quadratic form.

The fact that QUBO is unconstrained is circumvented by the addition of penalty terms \cite{stollenwerk2020image,glover2018tutorial}, which are extra terms that raise the value of $E(\boldsymbol x)$ whenever $\boldsymbol x$ is not a feasible solution --- i.e. when $\boldsymbol x$ does not obey all constraints. Importantly, imperfections in the penalization and minimization may lead to infeasible solutions.

Because of the reduced number of available qubits, we have devised a denser encoding of our problem into binary variables, mainly related to the representation of mono and stereo photos. In Section \ref{sec:classical} we have presented the classical formulation using our first encoding, which we refer to as \texttt{4cam}. It is a straightforward and fixed-length representation of the binary variables, whose allocation depends only on the total number of photos requested. In the denser encoding, which we call \texttt{3cam}, if requested photo $i$ is mono, there are three variables $x_{i,1}$, $x_{i,2}$, $x_{i,3}$, whereas if it is stereo there exists a single variable $x_i$. The advantages of the \texttt{3cam} formulation are the reduction in the number of qubits necessary and also in the number of constraints, since Eqs. (\ref{const:mono}, \ref{const:stereo}) no longer need to be imposed.

Finally, the polynomial to be minimized must be of quadratic form, a limitation particularly relevant for the penalty terms relative to ternary constraints \eqref{const:ternary}. These require the introduction of additional binary variables (``slack'' variables), which is another source of overhead in the amount of qubits needed. Let us momentarily simplify the notation of inequality \eqref{const:ternary} to $x_{p} + x_{q} + x_{r} \leq 2$ for cleanness. For \texttt{4cam} we write the corresponding penalty term directly in quadratic form, introducing two slack variables $s_1, s_2$ for each ternary constraint \cite[Sec 5]{glover2018tutorial}:
\begin{equation}
 P(x_p+x_q+x_r-2+s_1+2s_2)^2  \ ,
\label{eq:penalty4cam}
\end{equation}
where $P>0$ is a parameter. For \texttt{3cam}, in the spirit of optimizing this formulation as much as possible, we take a more involved approach: initially we write a cubic penalty term $P \ x_{p} x_{q} x_{r}$ and then reduce it to quadratic following Boros et al. \cite[Sec 4.4]{Boros2002}: a single slack variable $s_1$ replaces the pair $x_{q} x_{r}$, and an extra term is added (parentheses below),
\begin{equation}
P \ x_p s_1 + P ( x_qx_r-2x_qs_1-2x_rs_1+3s_1) \ .
\label{eq:penalty3cam}
\end{equation}
Advantages of the latter method are fewer terms in Eq.~\eqref{eq:penalty3cam} than in Eq.~\eqref{eq:penalty4cam} after expansion, and avoiding the introduction of a second slack variable per constraint. Additionally, if the same pair $x_qx_r$ appears in many constraints, the same slack variable replaces it in all terms, with the same extra term added.

\begin{figure}
    \centering
    \includegraphics[width=0.7\textwidth]{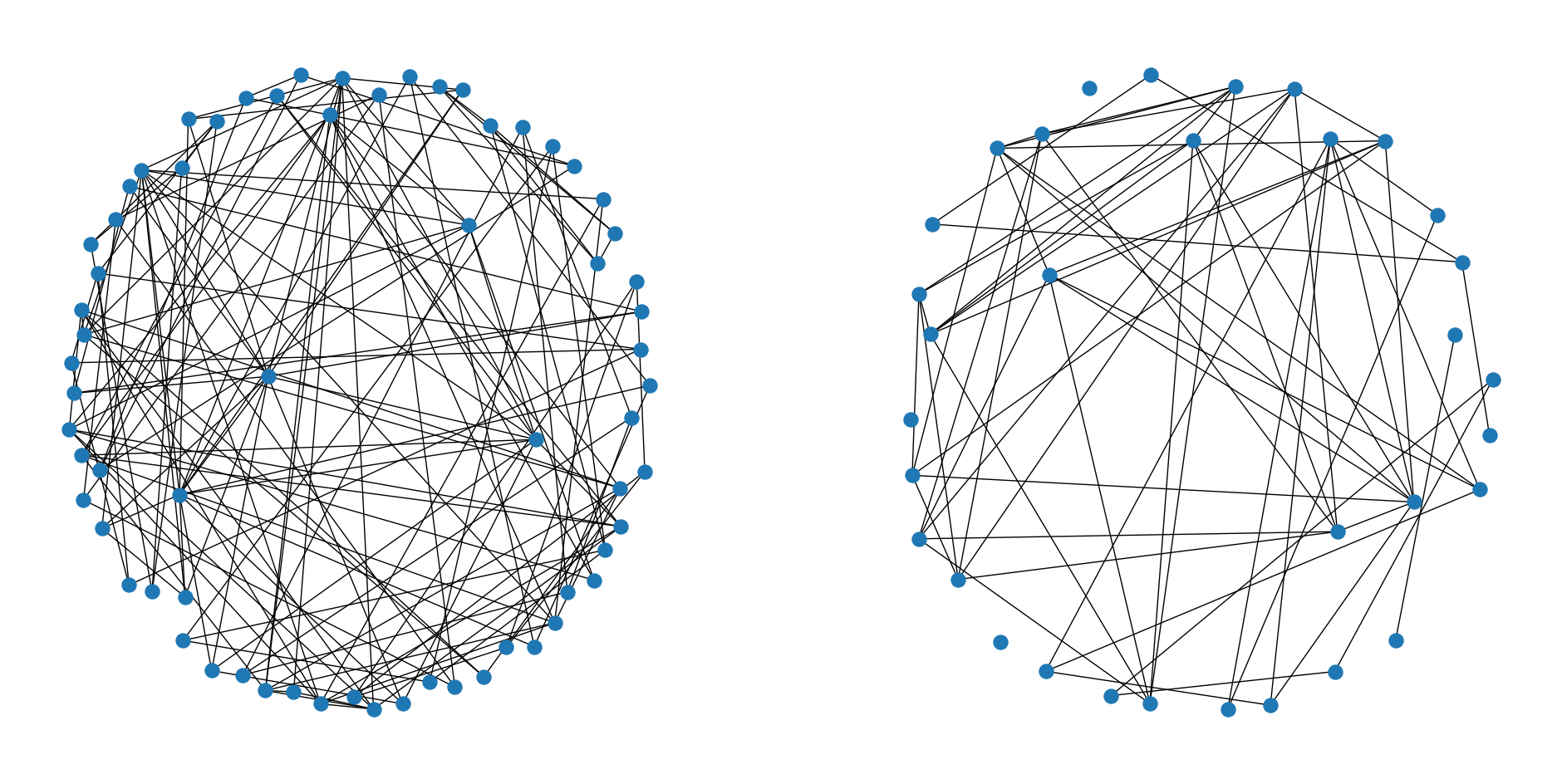}
    \caption{Graph representation of problem instance 15 (see the details in Table \ref{tab:instances}) obtained from using the QUBO $Q$ matrices as adjacency matrices. The one on the left corresponds to the \texttt{4cam} formulation with binary expansion encoding (Eq.\ref{eq:penalty4cam}) and the one on the right to the \texttt{3cam} formulation with Boros encoding (Eq.\ref{eq:penalty3cam}).
    For clarity, the self-loops (one for each node, corresponding to the diagonal of the matrix) have been omitted.}
    \label{fig:i15comparison}
\end{figure}

Figure \ref{fig:i15comparison} graphically shows the compression achieved with the formulation and encoding for instance 15. Notice that the representations include all variables, original and slack, and all necessary penalty terms. See Table \ref{tab:instances} for a full breakdown of all instances.

\section{Experimentation}\label{sec:exp}
This section is devoted to describe the experimentation carried out. First, we detail the benchmark of instances employed for testing the formulations detailed in the previous section. After that, we delve on the experimental setup and the analysis of the results obtained.

\subsection{Benchmark}
To conduct our study, the well-known SPOT5 benchmark dataset introduced in \cite{bensana1999earth} have been used, which is composed of 21 simulated instances of image-acquisition planning problems from the SPOT5 earth observation satellite of the French space agency. Among the 21 instances, 8 were discarded as they have capacity constraints, which consideration is out of the scope of this research. This dataset has certain characteristics that make it suitable for consideration in our study: it is open, heterogeneous, real-world oriented and widely used by the scientific community.

However, the large size of many of the instances is a limiting factor to asses the performance of the QPUs. To mitigate this, we have implemented a Python script for the automatic generation of reduced instances. This script, coined \texttt{InstanceReductor}, takes as input an existing instance of the SPOT5 dataset and the desired size for the newly generated instance. Then, the \texttt{InstanceReductor} generates a new file by randomly reducing the data in order to contemplate the number of requests introduced as input.

Overall, 20 different instances have been employed in the experimentation designed in this study. 14 of them are part of SPOT5, and their sizes range from 8 to 348 requests. The 6 remaining cases have been generated by \texttt{InstanceReductor}, and they consider a number of requests between 15 and 40. We summarize in Table \ref{tab:instances} the characteristics of each instance considered in this paper. Lastly, and aiming to enhance the reproducibility of this study, both generated cases and \texttt{InstanceReductor} are openly available in \cite{dataset}.

\begin{table}[!t]
\caption{Main characteristics of the 20 used instances, ordered by increasing number of requests. For each instance we depict the number of total and stereo requests, the amount of total and ternary constraints as well as the number of linear (L4cam, L3cam) and quadratic (Q4cam, Q3cam) terms for the two QUBO formulations. Shaded instances are generated by \texttt{InstanceReductor}.}
\label{tab:instances}
 \centering
 \resizebox{0.75\columnwidth}{!}{
     \begin{tabular}{l cc cc | cc | cc}
      \toprule
      ID & Requests & Stereo & Constraints & Ternary & L4cam & Q4cam & L3cam & Q3cam \\
      \midrule      
      \texttt{8}                  & 8   & 4   & 7    & 0    & 32    & 65    & 16   & 29    \\ 
      \rowcolor{Gray} \texttt{15} & 15  & 6   & 14   & 3    & 66    & 147   & 33   & 62    \\
      \rowcolor{Gray} \texttt{20} & 20  & 12  & 23   & 0    & 80    & 171   & 36   & 75    \\ 
      \rowcolor{Gray} \texttt{25} & 25  & 5   & 34   & 0    & 100   & 213   & 39   & 84    \\ 
      \rowcolor{Gray} \texttt{30} & 30  & 10  & 69   & 0    & 120   & 320   & 52   & 173   \\ 
      \rowcolor{Gray} \texttt{35} & 35  & 23  & 64   & 4    & 148   & 331   & 52   & 119   \\ 
      \rowcolor{Gray} \texttt{40} & 40  & 27  & 139  & 0    & 160   & 431   & 56   & 215   \\ 
      \texttt{54}                 & 67  & 35  & 204  & 23   & 314   & 997   & 140  & 544   \\ 
      \texttt{29}                 & 82  & 29  & 380  & 0    & 328   & 1102  & 120  & 667   \\ 
      \texttt{404}                & 100 & 63  & 610  & 18   & 436   & 1672  & 176  & 1078  \\ 
      \texttt{503}                & 143 & 78  & 492  & 86   & 744   & 2294  & 310  & 1118  \\ 
      \texttt{42}                 & 190 & 19  & 1204 & 64   & 888   & 3438  & 315  & 2026  \\       
      \texttt{408}                & 200 & 120 & 2032 & 389  & 1578  & 6870  & 489  & 3624  \\
      \texttt{28}                 & 230 & 35  & 4996 & 590  & 2100  & 12139 & 512  & 7501  \\ 
      \texttt{505}                & 240 & 119 & 2002 & 526  & 2012  & 8280  & 694  & 4226  \\ 
      \texttt{412}                & 300 & 160 & 4048 & 389  & 1978  & 11495 & 705  & 7823  \\ 
      \texttt{5}                  & 309 & 46  & 5312 & 367  & 1970  & 18787 & 1052 & 15828 \\
      \texttt{507}                & 311 & 163 & 5421 & 2293 & 9246  & 24762 & 1053 & 9641  \\ 
      \texttt{509}                & 348 & 178 & 8276 & 3927 & 9246  & 39418 & 1431 & 14619 \\
      \texttt{414}                & 364 & 182 & 9744 & 4719 & 10894 & 46827 & 1643 & 17429 \\ 
      \bottomrule        
     \end{tabular}
}
\end{table}

\subsection{Setup and Results}
To conduct the experiments on quantum hardware, we have used our two QUBO formulations (\texttt{4cam} and \texttt{3cam}) on the 20 instances detailed in Table \ref{tab:instances}. Additionally, for each instance and encoding, we tested four different D-Wave solvers (\texttt{2000Q}, \texttt{Advantage}, \texttt{Advantage2} and \texttt{HybridBQM}, where the latter stands for \texttt{hybrid\_binary\_quadratic\_model\_version2}). To account for the probabilistic nature of the solvers, we run each combination 5 times. For the three completely quantum solvers we have left all parameters as default except the number of runs, which we have set at 2000 following the advice of D-Wave. For the hybrid solver, we adopt all the default parameters. Lastly, the value of all penalty coefficients $P$ of each instance has been set to one plus the maximum possible value that the objective function can reach. In this regard, refining the choice of $P$ could be further investigated as it can severely affect the feasibility of the solutions obtained.

Additionally, in order to use it as a reference for the quantum experiments, we have first implemented the classical \texttt{4cam} formulation described in Section \ref{sec:classical} and solved all the considered instances with \texttt{Google OR-Tools}, checking that our results coincide with the known optima.

\begin{figure}
    \centering
    \includegraphics[width=\textwidth]{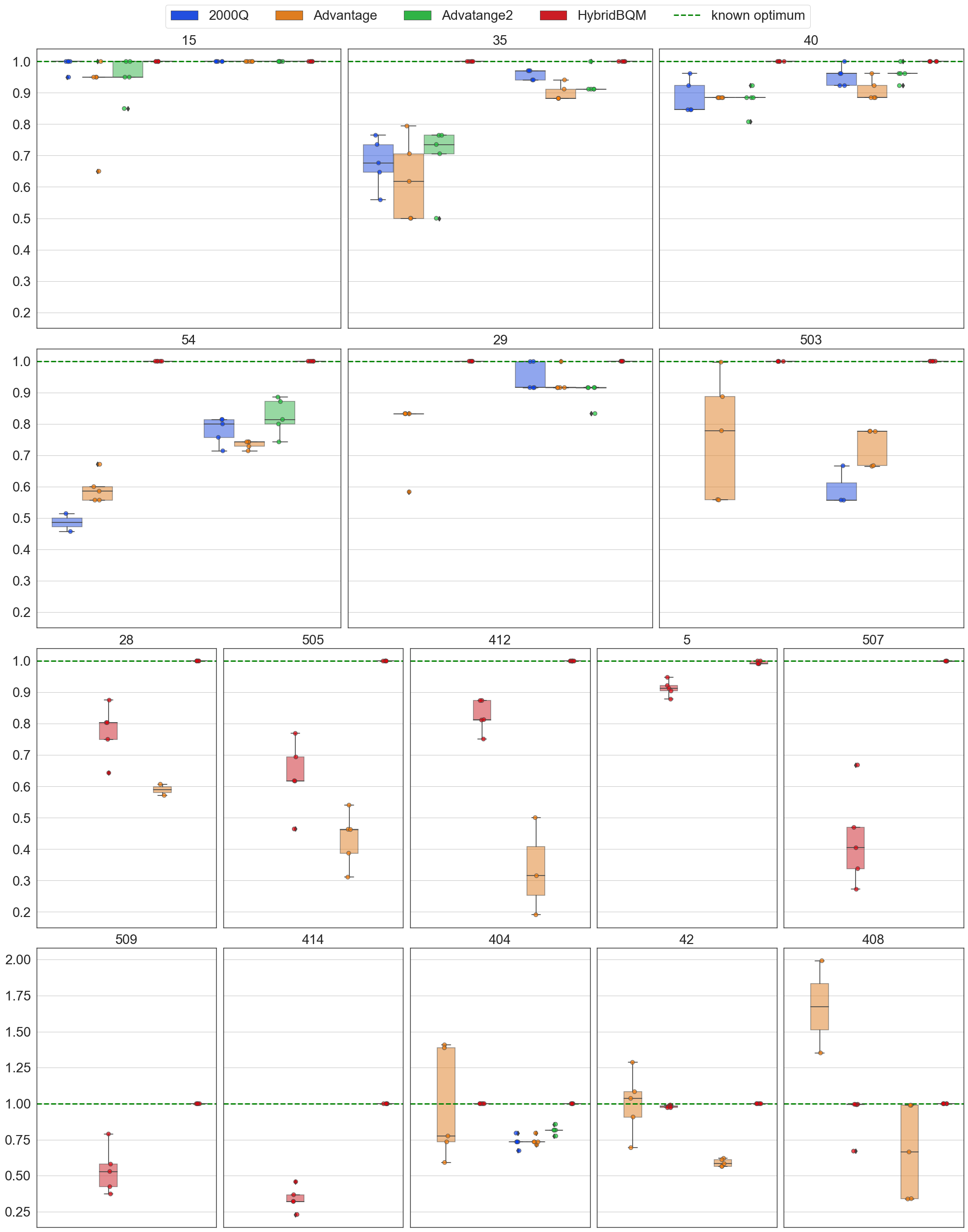}
    \caption{Box and jitter plots of the experimental results. Each subplot shows the execution data for one instance normalized to its optimum value. The dashed green line represents the optimal solution and the colors encode the solvers. The $x$ axis is split between the two formulations, \texttt{4cam} on the left and \texttt{3cam} on the right. The first three rows of the plot share the same $y$ axis, while the last row does not due to the broken constraints in instances \texttt{404, 42} and \texttt{408}. Instances \texttt{8, 20, 25} and \texttt{30}, with perfect performance up to 2 decimals in all cases, have been omitted. Elsewhere, when no data is shown for a given solver and instance, no embedding was found for it in any of the 5 runs.}
    \label{fig:all_results}
\end{figure}

Table \ref{tab:solutions} depicts the average results and standard deviations reached by \texttt{2000Q}, \texttt{Advantage}, \texttt{Advantage2} and \texttt{HybridBQM}, as well as by the classical solver. In Fig. \ref{fig:all_results} we show the detailed results of our experiments split by instance and formulation. Together with Table \ref{tab:solutions}, we can see that the better-performing formulation across almost all instances and solvers is \texttt{3cam}. Also, with \texttt{3cam} we obtain solutions for certain combinations of instance and solver that are untreatable with \texttt{4cam}. This is so because it is much more efficient in terms of variables and connections to be used for modelling the problem. Additionally, the best-performing solver is the \texttt{HybridBQM}.

\begin{table}[!t]
\caption{Results for the considered instances by encoding and solver. Each instance was run 5 times and the values reported are the mean $\pm$ standard deviation of the objective function value. Marked in bold are the best-performing results for each problem instance ignoring the hybrid solver. Results marked with * are those for which an embedding was not found in at least one of the runs while the ones with no numerical values are the ones for which no embedding was found in any of the 5 attempts. Instances \texttt{404}, \texttt{42} and \texttt{408} with the \texttt{4cam} formulation and \texttt{Advantage} QPU had some unfeasible solutions, which were removed when computing the results shown in this Table.}
\label{tab:solutions}
\centering
\resizebox{1\columnwidth}{!}{
\begin{tabular}{l cccc cccc}
\toprule
 & \multicolumn{4}{c}{\texttt{4cam}} & \multicolumn{4}{c}{\texttt{3cam}} \\
\cmidrule(lr){2-5} \cmidrule(lr){6-9}
ID & \texttt{2000Q} & \texttt{Advantage} & \texttt{Advantage2} & \texttt{HybridBQM} & \texttt{2000Q} & \texttt{Advantage} & \texttt{Advantage2} & \texttt{HybridBQM} \\

\midrule

\texttt{8}   & \bftab1.0 $\pm$ 0.0    & \bftab1.0 $\pm$ 0.0   & \bftab1.0 $\pm$ 0.0   & 1.0 $\pm$ 0.0   & \bftab1.0 $\pm$ 0.0    & \bftab1.0 $\pm$ 0.0    & \bftab1.0 $\pm$ 0.0    & 1.0 $\pm$ 0.0  \\
\texttt{15}  & 0.99 $\pm$ 0.02        & 0.90 $\pm$ 0.13        & 0.95 $\pm$ 0.05       & 1.0 $\pm$ 0.0   & \bftab1.0 $\pm$ 0.0    & \bftab1.0 $\pm$ 0.0    & \bftab1.0 $\pm$ 0.0    & 1.0 $\pm$ 0.0  \\
\texttt{20}  & \bftab1.0 $\pm$ 0.0    & \bftab1.0 $\pm$ 0.0   & \bftab1.0 $\pm$ 0.0   & 1.0 $\pm$ 0.0   & \bftab1.0 $\pm$ 0.0    & \bftab1.0 $\pm$ 0.0    & \bftab1.0 $\pm$ 0.0    & 1.0 $\pm$ 0.0  \\
\texttt{25}  & \bftab1.0 $\pm$ 0.0    & \bftab1.0 $\pm$ 0.0   & \bftab1.0 $\pm$ 0.0   & 1.0 $\pm$ 0.0   & \bftab1.0 $\pm$ 0.0    & \bftab1.0 $\pm$ 0.0    & \bftab1.0 $\pm$ 0.0    & 1.0 $\pm$ 0.0  \\
\texttt{30}  & \bftab1.0 $\pm$ 0.0    & \bftab1.0 $\pm$ 0.0   & \bftab1.0 $\pm$ 0.0   & 1.0 $\pm$ 0.0   & \bftab1.0 $\pm$ 0.0    & \bftab1.0 $\pm$ 0.0    & \bftab1.0 $\pm$ 0.0    & 1.0 $\pm$ 0.0  \\
\texttt{35}  & 0.68 $\pm$ 0.07        & 0.62 $\pm$ 0.12       & 0.69 $\pm$ 0.10        & 1.0 $\pm$ 0.0   & \bftab0.96 $\pm$ 0.01  & 0.90 $\pm$ 0.02         & 0.93 $\pm$ 0.04        & 1.0 $\pm$ 0.0  \\
\texttt{40}  & 0.88 $\pm$ 0.05        & 0.88 $\pm$ 0.00        & 0.88 $\pm$ 0.04       & 1.0 $\pm$ 0.0   & 0.95 $\pm$ 0.03        & 0.91 $\pm$ 0.03        & \bftab 0.96 $\pm$ 0.02 & 1.0 $\pm$ 0.0  \\
\texttt{54}  & 0.49 $\pm$ 0.03*       & 0.59 $\pm$ 0.04       & --                    & 1.0 $\pm$ 0.0   & 0.78 $\pm$ 0.04        & 0.73 $\pm$ 0.01        & \bftab0.82 $\pm$ 0.05  & 1.0 $\pm$ 0.0  \\
\texttt{29}  & --                     & 0.78 $\pm$ 0.10        & --                    & 1.0 $\pm$ 0.0   & \bftab0.95 $\pm$ 0.04  & 0.93 $\pm$ 0.03        & 0.90 $\pm$ 0.03         & 1.0 $\pm$ 0.0  \\
\texttt{404} & --                     & 0.70 $\pm$ 0.08        & --                    & 1.0 $\pm$ 0.0   & 0.73 $\pm$ 0.04        & 0.74 $\pm$ 0.03        & \bftab0.82 $\pm$ 0.03  & 1.0 $\pm$ 0.0  \\
\texttt{503} & --                     & \bftab0.76 $\pm$ 0.18 & --                    & 1.0 $\pm$ 0.0   & 0.59 $\pm$ 0.05*       & 0.73 $\pm$ 0.05        & --                     & 1.0 $\pm$ 0.0  \\
\texttt{42}  & --                     & \bftab0.80 $\pm$ 0.11  & --                    & 0.98 $\pm$ 0.01 & --                     & 0.59 $\pm$ 0.02        & --                     & 1.0 $\pm$ 0.0  \\
\texttt{408} & --                     & --                    & --                    & 0.93 $\pm$ 0.13 & --                     & \bftab0.66 $\pm$ 0.29  & --                     & 1.0 $\pm$ 0.0  \\
\texttt{28}  & --                     & --                    & --                    & 0.78 $\pm$ 0.08 & --                     & \bftab0.59 $\pm$ 0.02* & --                     & 1.0 $\pm$ 0.0  \\
\texttt{505} & --                     & --                    & --                    & 0.63 $\pm$ 0.10  & --                     & \bftab0.43 $\pm$ 0.08  & --                     & 1.0 $\pm$ 0.0  \\
\texttt{412} & --                     & --                    & --                    & 0.82 $\pm$ 0.05 & --                     & \bftab0.34 $\pm$ 0.13* & --                     & 1.0 $\pm$ 0.0  \\
\texttt{5}   & --                     & --                    & --                    & 0.91 $\pm$ 0.02 & --                     & --                     & --                     & 0.99 $\pm$ 0.00 \\
\texttt{507} & --                     & --                    & --                    & 0.43 $\pm$ 0.14 & --                     & --                     & --                     & 1.0 $\pm$ 0.0  \\
\texttt{509} & --                     & --                    & --                    & 0.54 $\pm$ 0.15 & --                     & --                     & --                     & 1.0 $\pm$ 0.0  \\
\texttt{414} & --                     & --                    & --                    & 0.34 $\pm$ 0.07 & --                     & --                     & --                     & 1.0 $\pm$ 0.0  \\
\bottomrule
\end{tabular}
}
\end{table}

If we turn our attention to the purely quantum solvers, it is interesting that \texttt{Advantage2}, although being still a prototype at the time of writing this article, has the edge for some of the smaller instances where an embedding can be found. For larger instances, the \texttt{Advantage} QPU is the most reliable solver, which manages to obtain results (albeit not necessarily outstanding ones) up until instance \texttt{412}, with 300 requests.

Furthermore, \texttt{2000Q} cannot handle instances larger than \texttt{503} (143 requests), where the limit of \texttt{Advantage2} is at instance \texttt{404} (100 requests). The evolution of the solvers seems to be clear, and we can expect that when the final version of the \texttt{Advantage2} QPU is launched, we will be able to solve even larger problems with greater precision. Finally, an important note is that in instances \texttt{404}, \texttt{42} and \texttt{408} there are some solutions above the optimum value, which means that some constraints were broken in the solving process. This was likely due to insufficiently large penalty values, which highlights the importance of choosing them correctly.

\section{Conclusions and Further Work} \label{sec:conclusions}
In this paper we have experimentally assessed the performance of D-Wave's three pure quantum annealers and one hybrid solver for the SIASP using two distinct formulation approaches. We have resorted to a realistic benchmark dataset and established how the quality of the solutions degrades with problem size, imposing practical limits on instances that can currently be solved effectively. 

Our results show that an efficient formulation allows to solve larger problem instances with better accuracy. This fact is key for success, which makes it a promising avenue for further research. The parameterization also influences the quality of the solutions, leading us to believe that a more exhaustive tuning of the problem penalty values as well as the solver parameters such as number of reads, chain strength, annealing time, etc. could bring us better performances overall.

On another note, future research could be focused on extending the problem to consider capacity constraints or multiple satellites, which would make it more appealing for industrial applications. Finally a study of the problem from the perspective of gate-based quantum computers, for example by means of variational quantum algorithms such as the quantum approximate optimization algorithm, would also be of significant interest.

\section*{Acknowledgments}
This work was supported by the Government of Spain (Misiones CUCO Grant MIG-20211005, FIS2020-TRANQI and Severo Ochoa CEX2019-000910-S), Fundació Cellex, Fundació Mir-Puig, Generalitat de Catalunya (CERCA program), and by the European Research Council ERC AdG CERQUTE.

\bibliographystyle{splncs04}
\bibliography{biblio.bib}

\begin{thebibliography}{10}
\providecommand{\url}[1]{\texttt{#1}}
\providecommand{\urlprefix}{URL }
\providecommand{\doi}[1]{https://doi.org/#1}

\bibitem{barkaoui2020new}
Barkaoui, M., Berger, J.: A new hybrid genetic algorithm for the collection
  scheduling problem for a satellite constellation. Journal of the Operational
  Research Society  \textbf{71}(9),  1390--1410 (2020)

\bibitem{bensana1999earth}
Bensana, E., Lemaitre, M., Verfaillie, G.: Earth observation satellite
  management. Constraints  \textbf{4}(3),  293--299 (1999)

\bibitem{bianchessi2007heuristic}
Bianchessi, N., Cordeau, J.F., Desrosiers, J., Laporte, G., Raymond, V.: A
  heuristic for the multi-satellite, multi-orbit and multi-user management of
  earth observation satellites. European Journal of Operational Research
  \textbf{177}(2),  750--762 (2007)

\bibitem{Boros2002}
Boros, E., Hammer, P.L.: Pseudo-boolean optimization. Discrete Applied
  Mathematics  \textbf{123}(1),  155--225 (2002).
  \doi{https://doi.org/10.1016/S0166-218X(01)00341-9},
  \url{https://www.sciencedirect.com/science/article/pii/S0166218X01003419}

\bibitem{Choi2011}
Choi, V.: Minor-embedding in adiabatic quantum computation: {{II}}.
  {{Minor-universal}} graph design. Quantum Information Processing
  \textbf{10}(3),  343--353 (Jun 2011). \doi{10.1007/s11128-010-0200-3}

\bibitem{DW-Adv2p11-Manual}
D-Wave Systems, Burnaby, Canada: QPU-Specific Physical Properties:
  Advantage2\_prototype1.1 (User Manual) (2022),
  \url{https://docs.dwavesys.com/docs/latest/\_downloads/08c75269a89583c35e421c45c35437eb/09-1275A-B\_QPU\_Properties\_Advantage2\_prototype1\_1.pdf}

\bibitem{DW-Adv61-Manual}
D-Wave Systems, Burnaby, Canada: QPU-Specific Physical Properties:
  Advantage\_system6.1 (User Manual) (2022),
  \url{https://docs.dwavesys.com/docs/latest/\_downloads/1bfa16c9915114bdf8a37b14713c8953/09-1272A-A\_QPU\_Properties\_Advantage\_system6\_1.pdf}

\bibitem{DW-2000Q-Manual}
D-Wave Systems, Burnaby, Canada: QPU-Specific Physical Properties: DW\_2000Q\_6
  (User Manual) (2022),
  \url{https://docs.dwavesys.com/docs/latest/\_downloads/50b8fa700f78e5d5c4c3208e0a8377c9/09-1215A-D_QPU_Properties_DW_2000Q_6.pdf}

\bibitem{glover2018tutorial}
Glover, F., Kochenberger, G., Du, Y.: Quantum {{Bridge Analytics I}}: A
  tutorial on formulating and using {{QUBO}} models. 4OR  \textbf{17}(4),
  335--371 (Dec 2019). \doi{10.1007/s10288-019-00424-y}

\bibitem{1399860}
Lin, W.C., Liao, D.Y.: A tabu search algorithm for satellite imaging
  scheduling. In: 2004 IEEE International Conference on Systems, Man and
  Cybernetics (IEEE Cat. No.04CH37583). vol.~2, pp. 1601--1606 vol.2 (2004).
  \doi{10.1109/ICSMC.2004.1399860}

\bibitem{luckow2021quantum}
Luckow, A., Klepsch, J., Pichlmeier, J.: Quantum computing: Towards industry
  reference problems. Digitale Welt  \textbf{5}(2),  38--45 (2021)

\bibitem{malladi2016satellite}
Malladi, K.T., Minic, S.M., Karapetyan, D., Punnen, A.P.: Satellite
  constellation image acquisition problem: A case study. Space Engineering:
  Modeling and Optimization with Case Studies pp. 177--197 (2016)

\bibitem{orus2019quantum}
Or{\'u}s, R., Mugel, S., Lizaso, E.: Quantum computing for finance: Overview
  and prospects. Reviews in Physics  \textbf{4},  100028 (2019)

\bibitem{dataset}
Osaba, E., Makarov, A., Taddei, M.M.: Benchmark dataset and instance generator
  for the satellite-image-acquisition scheduling problem.
  \url{http://dx.doi.org/10.17632/dzwvt4bz4j.1} (2023), online at Mendeley Data

\bibitem{osaba2022systematic}
Osaba, E., Villar-Rodriguez, E., Oregi, I.: A systematic literature review of
  quantum computing for routing problems. IEEE Access pp. 55805--55817 (2022)

\bibitem{ribeiro2010strong}
Ribeiro, G.M., Constantino, M.F., Lorena, L.A.N.: Strong formulation for the
  spot 5 daily photograph scheduling problem. Journal of combinatorial
  optimization  \textbf{20},  385--398 (2010)

\bibitem{stollenwerk2020image}
Stollenwerk, T., Michaud, V., Lobe, E., Picard, M., Basermann, A., Botter, T.:
  Image acquisition planning for earth observation satellites with a quantum
  annealer. arXiv preprint arXiv:2006.09724  (2020)

\bibitem{stollenwerk2021agile}
Stollenwerk, T., Michaud, V., Lobe, E., Picard, M., Basermann, A., Botter, T.:
  Agile earth observation satellite scheduling with a quantum annealer. IEEE
  Transactions on Aerospace and Electronic Systems  \textbf{57}(5),  3520--3528
  (2021)

\bibitem{vasquez2001logic}
Vasquez, M., Hao, J.K.: A “logic-constrained” knapsack formulation and a
  tabu algorithm for the daily photograph scheduling of an earth observation
  satellite. Computational optimization and applications  \textbf{20}(2),
  137--157 (2001)

\bibitem{vasquez2003upper}
Vasquez, M., Hao, J.K.: Upper bounds for the spot 5 daily photograph scheduling
  problem. Journal of Combinatorial Optimization  \textbf{7},  87--103 (2003)

\bibitem{wang2018exact}
Wang, J., Demeulemeester, E., Hu, X., Qiu, D., Liu, J.: Exact and heuristic
  scheduling algorithms for multiple earth observation satellites under
  uncertainties of clouds. IEEE Systems Journal  \textbf{13}(3),  3556--3567
  (2018)

\bibitem{wang2020agile}
Wang, X., Wu, G., Xing, L., Pedrycz, W.: Agile earth observation satellite
  scheduling over 20 years: Formulations, methods, and future directions. IEEE
  Systems Journal  \textbf{15}(3),  3881--3892 (2020)

\bibitem{zhang2018solving}
Zhang, Y., Hu, X., Zhu, W., Jin, P.: Solving the observing and downloading
  integrated scheduling problem of earth observation satellite with a quantum
  genetic algorithm. Journal of Systems Science and Information  \textbf{6}(5),
   399--420 (2018)

\bibitem{zhi2021variable}
Zhi, H., Liang, W., Han, P., Guo, Y., Li, C.: Variable observation duration
  scheduling problem for agile earth observation satellite based on quantum
  genetic algorithm. In: 2021 40th Chinese Control Conference (CCC). pp.
  1715--1720. IEEE (2021)

\end{thebibliography}
\end{document}